\documentclass[prl,aps,twocolumn,superscriptaddress,floatfix,citeautoscript,longbibliography]{revtex4-2}

\usepackage{graphicx,rotating,subfigure,amsmath,amsfonts,amssymb,delarray,color}

\usepackage{dsfont}
\usepackage[T1]{fontenc}
\usepackage{multirow}
\usepackage{comment}
\usepackage{hyperref}

\newcommand{\e}{\text{e}}

\def\nn{\nonumber}
\newcommand{\dr}[1]{\widetilde{#1}}

\newcommand{\ii}{\mathrm{i}}

\begin{document}

\title{Non-linear Transport by Bethe Bound States}
\author{Andrew Urichuk}
\email{urichuka@myumanitoba.ca}
\affiliation{Department of Physics and Astronomy and Manitoba Quantum Institute, University of Manitoba, Winnipeg R3T 2N2, Canada}
\affiliation{Fakult\"at f\"ur Mathematik und Naturwissenschaften, Bergische Universit\"at Wuppertal, 42097 Wuppertal, Germany}
\author{Andreas Kl\"umper}
\affiliation{Fakult\"at f\"ur Mathematik und Naturwissenschaften, Bergische Universit\"at Wuppertal, 42097 Wuppertal, Germany}
\author{Jesko Sirker}
\affiliation{Department of Physics and Astronomy and Manitoba Quantum Institute, University of Manitoba, Winnipeg R3T 2N2, Canada}

\begin{abstract}
We consider non-linear ballistic spin transport in the XXZ spin chain and derive an analytical result for the non-linear Drude weight $D^{(3)}$ at infinite temperatures. In contrast to the linear Drude weight $D^{(1)}$, we find that the result not only depends on anisotropy but also on the string length of the quasiparticles transporting the spin current. Our result provides further insights into transport by quasiparticles and raises questions about Luttinger liquid universality.
\end{abstract}
\maketitle

\paragraph{Introduction---$\!\!\!\!$}
Transport in metals is usually well described by a Boltzmann equation for long-lived quasiparticles \cite{Ziman_book,AshcroftMermin}. 
Standard transport theory though cannot necessarily be expected to hold for one-dimensional quantum systems. Here, Fermi liquid theory breaks down and the low-energy properties are instead described by Luttinger liquid theory which is based on collective excitations rather than quasiparticles \cite{GiamarchiBook}. For such systems, transport in the linear response regime can be studied using standard many-body techniques based on the Kubo formulas \cite{Mahan}. For a clean one-dimensional metal at zero temperature, one finds that the dc-conductivity has a zero-frequency Drude peak $D$ whose weight is determined entirely by the velocity of the collective excitations $v$ and the Luttinger parameter $K$ \cite{ShastrySutherland,ScalapinoWhite}. 

At finite temperatures, currents are always able to relax even in a clean one-dimensional lattice system as long as at least two non-commuting Umklapp scattering processes are present \cite{RoschAndrei}. This typically leads to diffusive transport \cite{SirkerPereira,SirkerPereira2,sirker_transport_2020}. An exception are one-dimensional integrable models which have an extensive number of local conservation laws that can protect a current from decaying completely \cite{zotos,zotos_tba_2017,KluemperSakai,Sirker_IJMPB,sirker_transport_2020}. Integrable models are not just a mathematical oddity: close realizations have also attracted considerable experimental attention \cite{MotoyamaEisaki,TakigawaMotoyama,ThurberHunt,MourigalEnderle,SologubenkoGianno,KinoshitaWenger,ParedesWidera}. In particular, good realizations of the spin-1/2 Heisenberg chain---which can also be viewed as a system of interacting spinless fermions---have recently been achieved in cold atomic gases and dynamical and transport properties have been studied extensively \cite{FukuharaKantian,FukuharaSchauss,HildFukuhara,jepsen_spin_2020,wei2021quantum}.

Gapless integrable quantum systems show an interesting duality: Thermodynamic quantities at low energies can be calculated essentially exactly within the Luttinger liquid framework if integrability is understood as a fine-tuning of the parameters in the field theory fixing the velocity of the collective excitations, the Luttinger parameter, and the amplitudes of irrelevant operators \cite{EggertAffleck92,Lukyanov,Kluemper_HB,Sirker_IJMPB}. At the same time, thermodynamic quantities can also be calculated using the thermodynamic Bethe ansatz (TBA) which deals with particles and their bound states. Thus, both a picture of collective excitations and of quasiparticles seem to be equally valid. 

For the transport properties of the anisotropic spin-1/2 Heisenberg (XXZ) chain, however, this no longer seems to be the case. It was recently demonstrated using the TBA formalism that the Drude weight of this model is fractal as a function of anisotropy not only at infinite  \cite{Prosen,ProsenIlievski} but also at low temperatures \cite{urichuk_analytical_2021}, contradicting 
Luttinger liquid theory \cite{SirkerPereira,SirkerPereira2,Sirker_IJMPB}. A similar fractal structure has also been obtained for the return probability following a quench from a domain-wall state~\cite{stephan_return_2017}. More generally, it has been argued that the non-equilibrium properties of integrable models can be understood using a Bethe-Boltzmann equation which includes densities for all conserved charges leading to a generalized hydrodynamical (GHD) description \cite{bulchandani_bethe-boltzmann_2018,doyon_drude_2017,castro-alvaredo_emergent_2016}. However, while the Drude weight is fractal for all finite temperatures, it strangely does not reveal the full particle content of the model: the type of magnon bound state carrying the current for a given anisotropy does not directly enter. 

In this letter we will extend the study of transport in the anisotropic spin-1/2 Heisenberg chain to the non-linear response regime. We will, in particular, study the most singular part of the non-linear response obtained when all relevant frequencies in a generalized conductivity are sent to zero, leading to the recently introduced notion of non-linear Drude weights \cite{oshikawa_quantum_2020,watanabe_general_2020}. These weights can be understood as a straightforward generalization of the Kohn formula \cite{Kohn,CastellaZotos}, relating the Drude weights to changes in the individual energy levels when a magnetic flux $\phi$ pierces a ring. Our main result is an exact formula for the non-linear Drude weight $D^{(3)}$ at infinite temperatures, demonstrating that in this response function the full particle content of the model is revealed. 

\paragraph{Non-linear Drude weights---$\!\!\!\!$}
It has been shown by Kohn \cite{Kohn} that the Drude weight at zero temperature can be understood as the response of the ground state energy to a magnetic flux through a ring. Alternatively, this can also be understood as the response to a twist in the boundary conditions. 
Later, this result was generalized to finite temperatures \cite{CastellaZotos}. In general, we can express all Drude weights as
\begin{equation}
\label{Drude}    
D^{(l)} = \frac{N^{l}}{Z}\sum_i \e^{-\beta E_i} \left.\frac{\partial^{l+1}E_i}{\partial \phi^{l+1}}\right|_{\phi=0} \, .
\end{equation}
Here $Z$ is the partition function, $N$ the number of sites, $E_i$ the energy levels, and $\beta=1/T$ the inverse temperature. The linear Drude weight $D^{(1)}$ is given by the curvature of the energy levels while non-linear Drude weights $(l>1)$ correspond to higher order derivatives. Note that because of the $\phi\to -\phi$ symmetry of the response, all Drude weights with $l$ even will vanish. The linear Drude weight can be obtained as $D^{(1)}=\lim_{t\to\infty}\lim_{N\to\infty}\frac{1}{2NT}\langle J(t)J(0)\rangle$ where $J(t)$ is the current operator at time $t$ and $\langle\cdots\rangle$ denotes the thermal average \cite{SirkerPereira2,sirker_transport_2020}. It thus describes the non-decaying, ballistic part of the current. In a similar way, the non-linear Drude weights can be understood as being given by more complicated multi-point current correlation functions \cite{watanabe_general_2020}.  

A simplification of the generalized Kohn formula \eqref{Drude} is obtained by noting that a one-dimensional system does not support a persistent current at zero flux in the thermodynamic limit. Therefore $\lim_{N\to\infty}\partial^l f/\partial \phi^l|_{\phi=0}=0$ where $f$ is the free energy density. This relation can be used to express $D^{(l)}$ by flux derivatives of order $l$ and lower only. A straightforward calculation shows, in particular, that 
\begin{eqnarray}
    \label{Drude2}
D^{(3)} &=& \frac{N^3}{Z}\sum_i \e^{-\beta E_i} \left.\frac{\partial^{4}E_i}{\partial \phi^{4}}\right|_{\phi=0} = \frac{N^3\beta}{Z}\sum_i \e^{-\beta E_i} \nonumber \\
&\times& \left\{\beta^2 \dot E_i^4 -6\beta \dot E_i^2\ddot E_i + 3\ddot E_i^2 + 4 \dot E_i \dddot E_i\right\} 
\end{eqnarray}
where the dots denote flux derivatives. This formula is also useful for numerical evaluations, see the Suppl.~Mat.

\paragraph{Model and non-interacting limit---$\!\!\!\!$}
In the following, we will calculate $D^{(3)}$ for the spin-1/2 XXZ chain
\begin{equation}
\label{Ham}
H\!=\!\frac{1}{4}\sum_{j=1}^N \left(\e^{-\ii \frac{\phi}{N}}\sigma^+_j\sigma^-_{j+1} +\e^{\ii \frac{\phi}{N}}\sigma^-_j\sigma^+_{j+1} +\Delta\sigma^z_j\sigma^z_{j+1}\right) 
\end{equation}
for anisotropies in the critical regime, $|\Delta|=|\cos(\gamma)| < 1$, with $\gamma = \frac{n \pi}{m}$ and $n , m \in \mathbb{Z}$. The model can be mapped to spinless fermions by a Jordan-Wigner transform with $\Delta$ then describing the strength of a nearest-neighbor interaction. For $\Delta=0$, the model is equivalent to free spinless fermions. Like the linear Drude weight $D^{(1)}$~\cite{ShastrySutherland}, the nonlinear Drude weights for the XXZ chain also exhibit a smooth dependence on the anisotropy at $T=0$ \cite{tanikawa_exact_2021}, however the nonlinear Drude weights $D^{(l)}$, $l>1$,  diverge if $\gamma < \frac{l-1}{l+3} \pi$. The cause being non-analytic flux corrections, which stem from Umklapp scattering. 

The $T>0$ linear Drude weight for the XXZ chain was first calculated in Ref.~\cite{zotos}. Here it was noted that it has the same structure as in the non-interacting case once the non-interacting particles are replaced by the appropriately dressed quasiparticles and their distribution functions. Such an equivalence was also later found for heat transport \cite{zotos_tba_2017}. The same quasiparticle structure is also underlying the GHD approach which has been successfully applied to integrable systems in the linear response regime and beyond \cite{castro-alvaredo_emergent_2016,bertini_transport_2016,doyon_drude_2017,bulchandani_solvable_2017,de_nardis_correlation_2021}. In particular, the analytical expression for the infinite temperature linear Drude weight first derived in Refs.~\cite{benz_finite_2005,Prosen,ProsenIlievski,PereiraPasquier} based on a set of quasilocal conserved charges has been reproduced in GHD \cite{collura_domain-wall_2020} and also directly from the TBA equations \cite{urichuk_spin_2019}. An extension to the GHD framework was proposed in~\cite{fava_hydrodynamic_2021} for numerically calculating nonlinear conductivities, which was then applied to the case of $\Delta>1$. 
At the moment there are, however, no non-zero temperature results for the non-linear Drude weights in the critical regime of the XXZ spin chain.

Let us first calculate $D^{(3)}$ in the non-interacting case, $\Delta=0$. In this case, the free energy density is given by $f=-N^{-1}T\sum_k\ln[1+\exp(-\beta\varepsilon_k)]$ with dispersion $\varepsilon_k=-\cos(k+\phi/N)$. It is now straightforward to take derivatives with respect to the flux $\phi$. Using again that $\partial^4 f/\partial\phi^4=0$ in the thermodynamic limit, we arrive at the following expression for the non-linear Drude weight 
\begin{eqnarray}
\label{Drude_D0}
D^{(3)} &=&N^3\sum_k n_k \varepsilon_k^{(4)} =N^3\beta\sum_k  n_k\bar n_k\left\{ 3 \ddot\varepsilon_k^2+4\dot\varepsilon_k\dddot{\varepsilon_k} \right. \nonumber \\
&+& \left. 6\beta (2n_k-1)\dot\varepsilon_k^2\ddot\varepsilon_k + \beta^2(1-6n_k\bar n_k)\dot\varepsilon_k^4\right\} \, .
\end{eqnarray}
Here $n_k=1/(\exp(\beta\varepsilon_k)+1)$ is the Dirac distribution and $\bar n_k = 1-n_k$. We will see below that the last expression is useful in the interacting case. Eq.~\eqref{Drude_D0} can be evaluated numerically for all finite temperatures and in closed form at zero and infinite temperatures, see the Suppl.~Mat.

\paragraph{The interacting case---$\!\!\!\!$}
From the Bethe ansatz equations, the quasiparticle momenta and phase shifts for the XXZ model can be obtained. Both can be characterized by so-called rapidities which arrange themselves, according to the string hypothesis, in regular patterns in the complex plane \cite{takahashi_thermodynamics_1999}. These strings describe bound states of magnons. Using the string hypothesis leads to the TBA equations which determine the Fermi weights $\vartheta_\alpha$ and dressed energies $\tilde\varepsilon_\alpha$ of the quasiparticles. Substituting these expressions into Eq.~\eqref{Drude_D0}, we will obtain an expression for $D^{(3)}$ in the interacting case which is proven in a more stringent manner in the Suppl.~Mat.

The flux $\phi$ enters the Hamiltonian \eqref{Ham} in the usual Peierls construction as a phase factor $\sim \exp(\pm \ii\phi/N)$. It is equivalent to a twist in the boundary conditions for the wave function $\Psi(l+N)=\e^{\ii\phi}\Psi(l)$. As such, it will not affect the properties of the system in the thermodynamic limit. We therefore need to calculate the finite-size corrections to the energy eigenvalues. Here we follow the approach in Refs.~\cite{zotos,fujimoto_exact_1998} for the linear Drude weight. 

The rapidities $\theta_\alpha$, which determine the quasimomenta $k_\alpha$, fulfill the Bethe ansatz equations
\begin{equation}
    \label{BA1}
\left(\frac{\sinh\frac{\gamma}{2}(\theta_\alpha+\ii)}{\sinh\frac{\gamma}{2}(\theta_\alpha-\ii)}\right)^N \!\!\!\! =-\e^{\ii\phi} \prod_{\alpha'=1}^M  \frac{\sinh\frac{\gamma}{2}(\theta_\alpha-\theta_{\alpha'}+2\ii)}{\sinh\frac{\gamma}{2}(\theta_\alpha-\theta_{\alpha'}-2\ii)}  
\end{equation}
for $\alpha=1,\cdots,M$ where $M$ is the number of spins down. The rapidities arrange themselves in regular patterns (strings) in the complex plane with the type of allowed strings $\alpha$ determined by the anisotropy $\Delta=\cos\gamma$. For a finite system of length $N$, we can now for each string type expand the rapidities $\theta_\alpha$ around the thermodynamic limit, resulting in 
\begin{equation}
    \label{BA2}
\theta_\alpha^N = \theta_\alpha^\infty +g_{1\alpha}\frac{\phi}{N}+g_{2\alpha}\left(\frac{\phi}{N}\right)^2+g_{3\alpha}\left(\frac{\phi}{N}\right)^3 +\cdots 
\, .  
\end{equation}
In the thermodynamic limit, we can obtain the densities of the $\theta_\alpha$ in terms of string densities $\rho_\alpha$ and hole densities $\rho_\alpha^h$, which can be combined into total densities $\rho^T_\alpha=\rho_\alpha + \rho^h_\alpha$ fulfilling the standard TBA integral equation
\begin{equation}
\label{TBA}
\rho^T_\alpha (\theta) = \frac{k'_\alpha(\theta)}{2\pi}+\sum_{\beta} \int d\lambda K_{\alpha\beta}\left(\theta - \lambda\right) \rho_\beta(\lambda) \,,
\end{equation}
where $K_{\alpha\beta}\left(\theta - \lambda\right)$ denotes the TBA kernel, see the Suppl.~Mat. for details.

From here, we can determine the finite size corrections $g_{j\alpha}$ in Eq.~\eqref{BA2}. In an expansion to leading order in inverse temperature $\beta$ we find, in particular, 
\begin{eqnarray}
\label{g-fcts}
\rho_\alpha^T g_{1\alpha} &=& \frac{\dr{q}_\alpha}{2\pi}=\frac{m}{4\pi} (\delta_{\alpha,m}+\delta_{\alpha,m-1}),\nn\\
\label{gs}
g_{n \alpha}  &=& \frac{1}{n}
g_{1\alpha}\partial_\theta g_{(n-1)\alpha} \,\, \text{ \, for \, } n>1 
\end{eqnarray}
with the dressed energy given by $\dr{\varepsilon}_\alpha = \frac{2\pi \sin\gamma}{\gamma} \sigma_\alpha \rho_\alpha^T$ and $\dr q_\alpha$ denoting the dressed spin. Furthermore, $\sigma_\alpha=\pm 1$ is a sign which depends on the type of string. We note that Eq.~\eqref{gs} reveals a very simple structure of the g-functions at infinite temperature: they are given by taking recursively higher derivatives of the inverse of the dressed energy. A dressed function $\dr f_\alpha$ is related to its bare quantity by the TBA integral relation
\begin{equation}
\label{dressing}
\dr f_\alpha (\theta)-\sum_{\beta} \int d\lambda\, K_{\alpha\beta}\left(\theta - \lambda\right) \vartheta_\beta(\lambda) \dr f_\beta (\lambda) = f_\alpha(\theta) \,.
\end{equation}
The Fermi weights $\vartheta_\alpha$ are given by $\vartheta_\alpha=(1+\eta_\alpha)^{-1}$ with $\eta_\alpha=\frac{\rho_\alpha^h}{\rho_\alpha}$. The infinite temperature case is particularly simple because dressed derivatives of the energy can directly be replaced by derivatives of the dressed energy $\dr\varepsilon_\alpha(\theta)$.
The derivatives of the latter with respect to the flux $\phi$ can straightforwardly be expressed by the finite size corrections $g_{j\alpha}$ using Eq.~\eqref{BA2}. We find, in particular\\
\begin{eqnarray}
\label{dressedE}
\frac{\partial\dr\varepsilon}{\partial \phi}\bigg\rvert_{\phi=0} &=& \partial_\theta \dr\varepsilon \frac{g_1}{N} \; ,\quad \frac{\partial^2\dr\varepsilon}{\partial \phi^2}\bigg\rvert_{\phi=0} = 2 \partial_\theta \dr\varepsilon \frac{g_2}{N^2} + \partial^2_\theta \dr\varepsilon \frac{g_1^2}{N^2} \nn \\
\frac{\partial^3\dr\varepsilon}{\partial \phi^3}\bigg\rvert_{\phi=0} &=& 6\partial_\theta \dr\varepsilon \frac{g_3}{N^3} + 6\partial^2_\theta \dr\varepsilon \frac{g_1 g_2}{N^3} + \partial^3_\theta \dr\varepsilon \frac{g_1^3}{N^3} \, .
\end{eqnarray}
Note that we have suppressed the subscript $\alpha$ for the string type. We are now in a position to express the Drude weight $D^{(3)}$ in Eq.~\eqref{Drude_D0} by the dressed energies and their distribution functions. To do so, we replace $\frac{1}{N}\sum_k \to \sum_\alpha\int d\theta \rho^T$, $n_k\to\vartheta_\alpha$, and $\partial_\phi\varepsilon_k \to  \partial_\phi\widetilde{\varepsilon}$.

\paragraph{High-temperature asymptotics---$\!\!\!\!$}
This then leads us to the following result for the high-temperature asymptotics
\begin{widetext}
\begin{equation}
    \label{D3main}
D^{(3)} \approx \beta\sum_\alpha\! \int\! d\theta \, \rho^T\vartheta (1-\vartheta)  \left\{(\partial_\theta\dr\varepsilon)^2\left(12g_{2}^2 + 24 g_{1}g_{3}\right) \!+ 36 (\partial_\theta\dr\varepsilon)(\partial^2_\theta\dr\varepsilon)g_{1}^2g_{2} \!+\!\left(3(\partial^2_\theta\dr\varepsilon)^2\!+4(\partial_\theta\dr\varepsilon)(\partial^3_\theta\dr\varepsilon)\right)g_{1}^4  \right\}  
\end{equation}
\end{widetext}
with the index $\alpha$ suppressed and the $g_{j\alpha}$ functions given by Eq.~\eqref{g-fcts}. Note that $g_{j\alpha}\sim (\dr{q}_\alpha)^j$  with the dressed spin given by $\dr q_\alpha=\frac{m}{2}(\delta_{\alpha,m}+\delta_{\alpha,m-1})$. Therefore only the last and second last strings will contribute to the sum over $\alpha$ in Eq.~\eqref{D3main} and both can be shown to give the same contribution. In order to carry out the explicit high temperature asymptotic calculation, the underlying $T$ and $Y$-systems~\cite{kuniba_t-systems_2011} for the XXZ chain are used to determine the last Fermi weight. The setup for this analysis is carried out in more detail in~\cite{urichuk_spin_2019,urichuk_analytical_2021} for the high and low-temperature asymptotics, respectively, and more details are given in the Suppl.~Mat. Here we want to demonstrate the method by first concentrating on the simple roots of unity case, $\Delta=\cos\gamma=\cos(\pi/m)$. In this case, it is straightforward to show that for the last string $\eta_\alpha = \frac{\rho_\alpha^h}{\rho_\alpha}=m-1$ for $\beta\to 0$. For the Fermi weights it follows that $\vartheta_\alpha(1-\vartheta_\alpha)=\eta_\alpha/(1+\eta_\alpha)^2=(m-1)/m^2$. To obtain the dressed energy $\dr\varepsilon_\alpha = \partial_\beta\ln\eta_\alpha$ to leading order, the first correction linear in $\beta$ to the $\eta_\alpha$ function is needed. One finds for $\alpha=m-1$
\begin{equation}
    \label{e_simple}
\dr\varepsilon_\alpha = \frac{m}{4(m-1)}\frac{\sin^2\gamma}{\cosh\frac{\gamma}{2}(\theta_\alpha+\ii)\cosh\frac{\gamma}{2}(\theta_\alpha-\ii)} \, .    
\end{equation}
Note that the prefactor $m/(m-1)=[m\vartheta_\alpha(1-\vartheta_\alpha)]^{-1}$, i.e., each factor of $\dr\varepsilon_\alpha$ brings in an inverse power of the distribution functions. We will see below that this explains why the particle content is hidden in the linear Drude weight while it is fully visible in the non-linear Drude weights. Finally, we note that the total density $\rho_\alpha^T = \rho_\alpha + \rho_\alpha^h$ can also be expressed by the dressed energy resulting in $\rho_\alpha^T = \sigma_\alpha \frac{\gamma}{2 \pi \sin\gamma}\dr\varepsilon_\alpha$. It is now straightforward to use the high temperature g-functions \eqref{g-fcts} to express the non-linear Drude weight \eqref{D3main} in terms of the Fermi weights $\vartheta_\alpha$ and dressed energies $\dr\varepsilon_\alpha$. The integral over the rapidities $\theta$ is then a convergent integral over hyperbolic functions which can be evaluated in closed form. The final result can be expressed as
\begin{eqnarray}
\label{eq:asymptoticFormula}
TD^{(l)} &=&  (-1)^{\frac{l-1}{2}}\frac{m^{2l-2}}{8} \left\{\vartheta (1-\vartheta) \right\}^{l-1} \\
&\times& \frac{\sin^2(\gamma)}{\sin^2\left(\frac{\pi}{m}\right)} \left[ 1 - \frac{(3l-2) m }{2\pi} \sin\left( \frac{2\pi}{m}\right) \right] \nonumber
\end{eqnarray}
with $l=3$ and we have suppressed the string index, which corresponds to either the last or second last string. In the Suppl.~Mat.~we explicitly show that the result above is valid for all anisotropies $\gamma=n\pi/m$ with $\gamma>\pi/3$. It is remarkable that the analytical expression for $D^{(3)}$ is quite similar to the one for the linear Drude weight $D^{(1)}$ \cite{ProsenIlievski} which is obtained from Eq.~\eqref{eq:asymptoticFormula} by setting $l=1$.

A couple of comments are in order: (i) In addition to the corrections $(\phi/N)^j$ with $j$ integer in Eq.~\eqref{BA2} there will also be corrections with non-integer powers. It is well known that the leading irrelevant operator for the XXZ chain stems from Umklapp scattering and has scaling dimension $2K$, where $K=\pi/(\pi-\gamma)$ is the Luttinger parameter (we adopt a definition where $K=1$ at the isotropic point, $\Delta=1$). Umklapp scattering leads to a $(\phi/N)^{4K-3}$ correction to the energies. To calculate the Drude weight $D^{(l)}$, we need an $l$-th derivative w.r.t.~the flux, see Eq.~\eqref{Drude2}. Therefore the expansion \eqref{BA2} is no longer valid if $4K-3<l$ which happens for anisotropies $\gamma/\pi < (l-1)/(l+3)$ \cite{tanikawa_fine_2021}. While our main result \eqref{eq:asymptoticFormula} remains finite in the entire critical regime $|\Delta|<1$, it is therefore only valid for $-1<\Delta<1/2$. (ii) While the result for $D^{(3)}$ in Eq.~\eqref{eq:asymptoticFormula} looks strikingly similar to the result for the linear Drude, there is a very important difference: while $D^{(3)}$ does explicitly contain the Fermi weight $\vartheta$, $D^{(1)}$ does not. Since $\vartheta\sim\bar\mu/m$ where $\bar\mu$ is the length of the last string, $D^{(3)}$ reflects the full bound-state particle content. For example, for $n=m-3$ the string length can take the two values $\bar\mu=(m\pm 1)/3$. $D^{(3)}$ is therefore even ,more fractal' than $D^{(1)}$. This statement can be made more precise: Let us consider approaching irrational anisotropies by sending $n,m\to\infty$ with $n/m$ fixed. We find $D^{(1)}\sim \frac{\beta}{12}\sin^2\gamma$, i.e., there is a continuous lower envelope function. On the other hand, $D^{(3)}\sim \frac{3\beta}{4}[\frac{m \bar\mu(m-\bar\mu)\sin\gamma}{\pi}]^2$ which means that the non-linear Drude weight is always diverging when approaching an irrational anisotropy value. 

From a phenomenological perspective, we expect that if there are stable quasiparticles which contribute to the transport, then their properties should enter the Drude weights explicitly. It is therefore surprising---an issue which does not seem to have been addressed explicitly so far---that the linear Drude weight $D^{(1)}$ does not depend on the string length $\bar\mu$ but rather only on $n,m$ which determine the anisotropy $\Delta=\cos(n\pi/m)$. Looking at the Drude weights $D^{(l)}$ more generally we see that $D^{(1)}\sim\int d\theta\, \vartheta(1-\vartheta)(\partial_\theta\dr\varepsilon)^2/\dr\varepsilon$ is an exceptional case. The dressed energy contains a factor $\dr\varepsilon\sim [\vartheta(1-\vartheta)]^{-1}$ so that the dependence on the Fermi weight $\vartheta\sim\bar\mu/m$ and thus the dependence on the string length $\bar\mu$ exactly cancels out in this case. 
More generally, we expect that the scaling with $l$ in the first line of Eq.~\eqref{eq:asymptoticFormula} also holds for $l>3$ because it is a direct consequence of the scaling of the dressed spin and dressed energy. It will be interesting to see if the structure in the second line also holds for $l>3$. We note that \eqref{eq:asymptoticFormula} for all odd  $l$ does give the correct result $TD^{(l)}=(-1)^{(l-1)/2}/8$ for $\Delta=0$.

\begin{figure}[!t]
	\includegraphics[width=0.95\columnwidth]{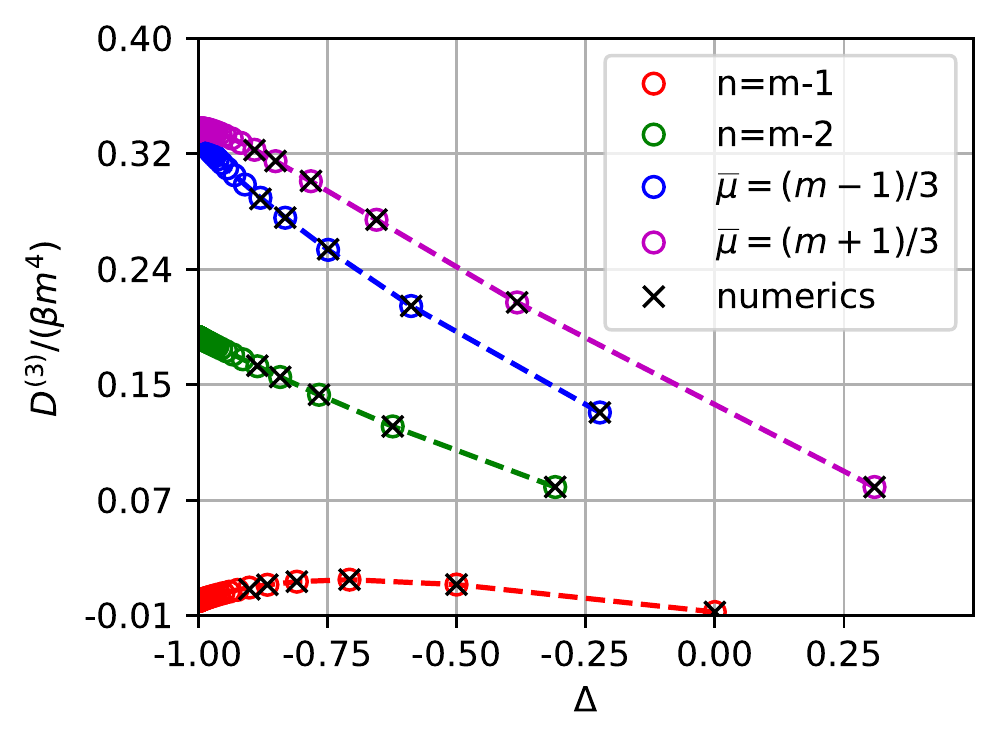}
	\caption{The non-linear Drude weight $TD^{(3)}$ scaled by $m^4$ to fit on a single scale. The lines are a guide for the eye with the note that \eqref{eq:asymptoticFormula} is only valid at rational points marked by an empty circle. Crosses are the result of numerically solving the non-linear Drude weight formula at $\beta = 10^{-4}$.} 
	\label{Fig2}
\end{figure}

$D^{(3)}$ as a function of anisotropy and string length is shown in Fig.~\ref{Fig2}. First, we note that instead of analytically solving for the g-functions as we have done in Eq.~\eqref{g-fcts}, we can also express them as a closed system of integral equations valid at all temperatures \cite{UrichukKluemperSirkerPrep}. This allows for a numerical evaluation of $D^{(3)}$. 
As shown in Fig.~\ref{Fig2}, the numerical results are consistent with the analytical formula \eqref{eq:asymptoticFormula} at $\beta\to 0$.
Note also that for the case $n=m-3$ there are two curves for the two different possible string lengths, clearly demonstrating that the current is transported by quasiparticles, the Bethe bound states.

\paragraph{Conclusions---$\!\!\!\!$}
We have shown that the non-linear transport properties of the XXZ chain are directly determined by the particle content of the theory by deriving an analytical result for the non-linear Drude weight $D^{(3)}$ at infinite temperatures. $D^{(3)}$ exhibits a nowhere continuous dependence on the anisotropy $\Delta=\cos(n\pi/m)$ just like the linear Drude weight $D^{(1)}$. However, in contrast to $D^{(1)}$ it also explicitly depends on the Bethe string length $\bar\mu$ and thus shows the full particle content of the theory. Our results shed further light on transport in metals which is typically understood using Boltzmann theory for long-lived quasiparticles. However, the validity of such an approach is usually difficult to examine except for free models. Here, integrable models can provide important benchmarks because their excitation spectra can contain complicated quasiparticles such as bound states which are exact, i.e., have infinite lifetime. Our results also shed further light on the dynamics of integrable systems by showing that Bethe strings are not merely a tool for calculations but appear indispensable. In particular, $D^{(3)}$ diverges when approaching irrational anisotropies where arbitrary string lengths are possible. This implies that Luttinger liquid theory cannot describe transport in such systems properly and raises the interesting question how these type of bound states can be incorporated in a field theory.

\acknowledgments
The authors acknowledge support by the German Research Foundation (DFG) via the Research Unit FOR2316. J.S.~acknowledges support by NSERC. 

\newpage
\appendix

\begin{widetext}

\section{Appendix A: High temperature g-functions}
In the main text, the Drude weight was found to result from the finite-size corrections caused by a flux $\phi$ through a ring. The $g$--functions characterize analytic corrections that distinguish the finite-size rapidity $\theta^N$ in the presence of a flux $\phi$ from the thermodynamic rapidity $\theta$ 
\begin{eqnarray}
\label{eq:rapExpand}
\theta^N_\alpha &=& \theta_\alpha + \frac{g_{1\alpha} \phi}{N} + \frac{g_{2\alpha} \phi^2}{N^2}+ \frac{g_{3\alpha} \phi^3}{N^3}+ \frac{g_{4\alpha} \phi^4}{N^4} + \mathcal{O}(N^{-5}).
\end{eqnarray}
Here $\alpha$ is the index characterizing the considered string. There is a subtle point here involving the possibility of non-analytic corrections of order $\mathcal{O}\left(N^{-4\gamma/(\pi-\gamma)}\right)$, which for $l=3$ are relevant for $\gamma \leq \frac{\pi}{3}$. Due to these non-analytic corrections, the following analysis is only valid for $\frac{\pi}{3}< \gamma< \pi$. To begin, we introduce the relevant portions of the standard thermodynamic Bethe ansatz (TBA). 

A set of rapidities $\{\lambda^a_\alpha\}$ uniquely characterize the quantum state. These Bethe roots are assumed to sort themselves into specific patterns known as strings indexed by string type $\alpha$. Key to the TBA relations is the definition of the scattering phase $\Theta_{\alpha\beta}$, defined as
\begin{eqnarray}
\Lambda_{\alpha}(\theta) &=& 2 \tan^{-1}\left(\frac{\tan \left(\ii\frac{\theta \gamma}{2}\right)}{\tanh\left(\ii\frac{\alpha \gamma}{2}\right)}\right) + 2\pi \left (\frac{\ii \gamma \theta + \pi}{2\pi}\right),\nn\\
\Theta_{\alpha \delta}(\theta) &=& \begin{cases}
\Lambda_{|\alpha-\delta|}+2\sum_{n=1}^{\delta-1} \Lambda_{|\alpha-\delta|+2n} +\Lambda_{\alpha+\delta} &\text{\, \,for\, \,} \alpha \neq \delta,\\
2\sum_{n=1}^{\alpha-1} \Lambda_{2n} +\Lambda_{2\alpha} &\text{\, \,for\, \,} \alpha = \delta.
\end{cases}
\end{eqnarray} 
 The standard TBA kernel is then given in terms of the scattering phase as
\begin{eqnarray}
K_{\alpha \delta}(\theta - \lambda_\delta) &=& \frac{\Theta'_{\alpha \delta}\left(\theta - \lambda_\delta\right)}{2\pi}.
\end{eqnarray} 

The strings behave as quasi-particles and can be used to rewrite the discrete Bethe roots in terms of continuous densities. In order to carry this out, we write out first the quantization condition for twisted boundary conditions and secondly introduce the counting function $Z_\alpha(\theta)$. The standard TBA relations are satisfied by a set of Bethe roots that are characterized by
\begin{eqnarray}
2\pi I^a_{\alpha} &=& N k_\alpha(\theta^a_{\alpha}) + \sum_{b, \delta} \Theta_{\alpha \delta}\left(\theta^a_{\alpha} - \lambda^b_{ \delta}\right)-\phi \mu_\alpha  ,\nn\\
Z_\alpha(\theta^a_\alpha) &=&  \frac{k_\alpha(\theta^a_\alpha)}{2\pi} + \sum_{b, \delta} \frac{\Theta_{\alpha \delta}\left(\theta^a_\alpha - \lambda^b_{ \delta}\right)}{2\pi N} -\frac{ \phi \mu_\alpha}{2\pi N},
\end{eqnarray}
where $\mu_\alpha$ is the string length of the string $\alpha$, $\phi$ is the applied twist angle, and $k_\alpha$ the bare momentum. These string patterns are regular enough to define the string particle density $\rho_\alpha$ and the hole density $\rho_\alpha^h$, which may be combined into the total density $\rho^T_\alpha=\rho_\alpha + \rho^h_\alpha$. 
By considering a small shift in the rapidity we can determine the density of Bethe roots $\lambda^\alpha_\alpha$. Considering the difference $Z_\alpha(\theta + d\theta) - Z_\alpha(\theta) $ then leads to the TBA equation defining the total density of the given string type
\begin{equation}
\label{Eq5}
Z_\alpha'(\theta)=\rho^T_\alpha(\theta) = \frac{k'_\alpha(\theta)}{2\pi}+\sum_{b, \delta} \frac{\Theta'_{\alpha \delta}\left(\theta - \lambda^b_{ \delta}\right)}{2\pi N} 
\to \frac{k'_\alpha(\theta)}{2\pi}+\sum_{\delta} \underbrace{\int d\lambda K_{\alpha\delta}\left(\theta - \lambda\right) \rho_\delta(\lambda)}_{K_{\alpha\delta}\star \rho_\delta}.
\end{equation}
The solutions may be characterized by $\eta_\alpha = \frac{\rho^h_\alpha}{\rho_\alpha}$, which may be identified with the underlying $Y$-system of the model. 
If we consider the finite-size rapidity $\theta^N_\alpha$ in the counting function, then the finite-size corrections can be determined by noting that 
\begin{eqnarray}
    Z_\alpha(\theta^N) &=&  \frac{k_\alpha(\theta^N)}{2\pi} + \sum_{b,\delta} \frac{\Theta_{\alpha\delta}\left(\theta^N - \lambda^b_{ \delta}\right)}{2\pi N} -\frac{ \phi \mu_\alpha}{2\pi N},\nn\\
    Z_\alpha(\theta)  &=&  \frac{k_\alpha(\theta)}{2\pi}+\sum_{\delta} \frac{\Theta_{\alpha\delta}}{2\pi} \star \rho_\delta+\frac{\phi}{N}\left(\frac{g_{1\alpha}  k_\alpha'(\theta)}{2\pi} +g_{1\alpha}\sum_{\delta} K_{\alpha\delta} \star \rho_\delta- \sum_{\delta} K_{\alpha\delta} \star (g_{1\delta}\rho_\delta)-\frac{  \mu_\alpha}{2\pi}\right).
\end{eqnarray}
 With the note that the $\mathcal{O}(N^{-1})$ contributions must independently vanish it follows that
\begin{eqnarray}
      && \frac{g_{1\alpha}  k_\alpha'(\theta)}{2\pi } +g_{1\alpha} \sum_{\delta} K_{\alpha\delta} \star \rho_\delta- \sum_{\delta} K_{\alpha\delta} \star (g_{1\delta}\rho_\delta)-\frac{ \mu_\alpha}{2\pi }=0,\nn\\
      &&g_{1\alpha} \rho^T_\alpha-\sum_{\delta} K_{\alpha\delta} \star (g_{1\delta}\rho_\delta)=\frac{ \mu_\alpha}{2\pi },\nn\\
      &&g_{1\alpha} \rho^T_\alpha=\frac{  \dr{q}_\alpha}{2\pi },\nn\\
      &&g_{1\alpha} = \sigma_\alpha\frac{\sin\gamma}{\gamma} \frac{\dr{q}_\alpha}{\dr \varepsilon_\alpha}.
\end{eqnarray}
In the second line we have used Eq.~\eqref{Eq5} and in the third line we have introduced the dressed spin $\dr{q}_\alpha=\mu_\alpha+2\pi\sum_\delta K_{\alpha\delta}\star (g_{1\delta}\rho_\delta)$. Furthermore, we have used that $\rho^T_\alpha=\frac{\sigma_\alpha}{2\pi}\frac{\gamma}{\sin\gamma}\dr{\varepsilon}_\alpha$ where $\dr{\varepsilon}_\alpha$ is the dressed energy defined as
\begin{eqnarray}
\partial_\beta \ln \eta_\alpha = \dr \varepsilon_\alpha = E_\alpha + \sum_\delta K_{\alpha\delta} \star (\vartheta_\delta \dr \varepsilon_\delta).
\end{eqnarray}

We can now use this first order contribution and our ansatz for the finite size corrections to write out the energy corrections in terms of higher order $g$-functions. At infinite temperatures, we note that the dressing commutes with differentiation, i.e. $\partial\dr{f} = \dr{\partial f} + \mathcal{O}(\beta)$, which we shall see is useful for determining the flux derivatives. There are two ways to build identities for the $g$-functions: the first is based on Taylor expanding the dressed energy around the thermodynamic rapidity 
\begin{eqnarray}
\label{Eq8}
\frac{\partial\dr\varepsilon_\alpha}{\partial \phi}\bigg\rvert_{\phi=0} &=& \partial_\theta \dr\varepsilon_\alpha \frac{g_{1\alpha}}{N} \; \nn\\
\frac{\partial^2\dr\varepsilon_\alpha}{\partial \phi^2}\bigg\rvert_{\phi=0} &=& 2 \partial_\theta \dr\varepsilon_\alpha \frac{g_{2\alpha}}{N^2} + \partial^2_\theta \dr\varepsilon_\alpha \frac{g_{1\alpha}^2}{N^2} \nn \\
\label{eq:dressedEapp}
\frac{\partial^3\dr\varepsilon_\alpha}{\partial \phi^3}\bigg\rvert_{\phi=0} &=& 6\partial_\theta \dr\varepsilon_\alpha \frac{g_{3\alpha}}{N^3} + 6\partial^2_\theta \dr\varepsilon_\alpha \frac{g_{1\alpha} g_{2\alpha}}{N^3} + \partial^3_\theta \dr\varepsilon_\alpha \frac{g_{1\alpha}^3}{N^3} \, .
\end{eqnarray}
The second method is to determine the second and third order flux derivatives by explicit differentiation, which leaves everything in terms of $g_{1\alpha}$. In the infinite temperature limit, we recall that differentiation commutes with the dressing operator and that the $g$-functions have no flux dependence to find
\begin{eqnarray}
\left(\frac{\partial}{\partial \phi}\frac{\partial\dr\varepsilon_\alpha}{\partial \phi}\right)\bigg\rvert_{\phi=0} &=& \left(\partial_\theta \partial_\phi\dr\varepsilon_\alpha\right)\bigg\rvert_{\phi=0} \frac{g_{1\alpha}}{N} = \partial_\theta\left(g_{1\alpha}\partial_\theta \dr\varepsilon_\alpha  \right)\frac{g_{1\alpha}}{N^2} = \left( \partial_\theta \dr\varepsilon_\alpha\right) \frac{g_{1\alpha}\left(\partial_\theta g_{1\alpha} \right)}{N^2}+\left(\partial^2_\theta \dr\varepsilon_\alpha\right)\frac{g_{1\alpha}^2}{N^2},\nn\\
\left(\frac{\partial}{\partial \phi}\frac{\partial^2\dr\varepsilon_\alpha}{\partial \phi^2}\right)\bigg\rvert_{\phi=0} &=& 2 \left(\partial_\theta \partial_\phi \dr\varepsilon_\alpha \right)\bigg\rvert_{\phi=0}\frac{g_{2\alpha}}{N^2} + \left(\partial^2_\theta \partial_\phi \dr\varepsilon_\alpha \right) \bigg\rvert_{\phi=0}\frac{g_{1\alpha}^2}{N^2}
= 2 \partial_\theta \left(g_{1\alpha} \partial_\theta \dr\varepsilon_\alpha \right)\frac{g_{2\alpha}}{N^3} + \partial^2_\theta\left( g_{1\alpha} \partial_\theta\dr\varepsilon_\alpha\right) \frac{g_{1\alpha}^2}{N^3}\nn\\
&=&\left( \partial_\theta \dr\varepsilon_\alpha \right)\frac{g_{1\alpha}^2\left( \partial_\theta^2 g_{1\alpha}\right) + 2g_{2\alpha}\partial_\theta g_{1\alpha}}{N^3} + \left( \partial^2_\theta \dr\varepsilon_\alpha \right)\frac{2g_{1\alpha} g_{2\alpha}+2 g_{1\alpha}^2 \left(\partial_\theta g_{1\alpha} \right)}{N^3} + \left(\partial^3_\theta \dr\varepsilon_\alpha\right) \frac{g_{1\alpha}^3}{N^3}.
\end{eqnarray}
By comparing the two forms for the flux derivatives we obtain
\begin{eqnarray}
g_{1\alpha} &=& \sigma_\alpha\frac{\sin(\gamma)}{\gamma} \frac{\dr{q}_\alpha}{\dr \varepsilon_\alpha},\nn \\
g_{2\alpha} &=& \frac{1}{2} g_{1\alpha}\left(\partial_\theta g_{1\alpha}\right)=\frac{1}{2} \frac{\sin^2\gamma}{\gamma^2} \frac{\dr{q}^2_\alpha}{\dr\varepsilon_\alpha}\partial_\theta\left( \frac{1}{\dr\varepsilon_\alpha}\right),\nn \\
g_{3\alpha} &=& \frac{1}{6} \left(g_{1\alpha}^2\left(\partial^2_\theta g_{1\alpha}\right)+2g_{2\alpha}\left(\partial_\theta g_{1\alpha}\right) \right) = \frac{1}{6} \sigma_\alpha \frac{\sin^3\gamma}{\gamma^3} \frac{\dr{q}^3_\alpha}{\dr\varepsilon_\alpha}\partial_\theta \left(\frac{1}{\dr\varepsilon_\alpha} \partial_\theta\left( \frac{1}{\dr\varepsilon_\alpha}\right)\right).
\end{eqnarray}

A clear pattern emerges, which allows us to write down the general form of the high-temperature $g$ functions. Rewritten in term of the dressed energy $\dr{\varepsilon}_\alpha$ on the RHS and the total density $\rho^T_\alpha$ on the LHS, we find
\begin{eqnarray}
\label{gfcts}
\rho_\alpha^T g_{1\alpha} &=& \frac{ \dr{q}_\alpha}{2\pi } = \frac{m }{4\pi} (\delta_{\alpha,m}+\delta_{\alpha,m-1}),\nn\\
\rho^T_\alpha g_{2\alpha } &=& \frac{1}{2}\frac{\sin(\gamma)}{2 \pi \gamma}\dr{q}_\alpha^2  \sigma_\alpha \partial_\theta\left(\frac{1}{\dr{\varepsilon}_\alpha}\right)+\mathcal{O}(\beta),\nn\\
\label{eq:highTgFunctions}
\rho_\alpha^T  g_{3\alpha } &=& \frac{1}{3!}\frac{\sin^2(\gamma)}{2 \pi \gamma^2} \dr{q}_\alpha^3 \partial_\theta\left( \frac{ 1}{\dr{\varepsilon}_\alpha} \partial_\theta\left(\frac{1}{\dr{\varepsilon}_\alpha}\right)\right)+\mathcal{O}(\beta).
\end{eqnarray}
These $g$-functions may then be written recursively as
\begin{eqnarray}
g_{1\alpha} &=& \sigma_\alpha\frac{\sin\gamma}{\gamma}\frac{\dr{q}}{\dr{\varepsilon}},\nn\\
\label{eq:g_functions}
g_{n \alpha}  &=& \frac{1}{n} g_{1\alpha}\partial_\theta g_{n-1,\alpha} \,\, \text{ \, for \, } n>1 .
\end{eqnarray}

\section{Appendix B: Infinite Temperature Formula}
In the main text, we found that the non-linear Drude weight $D^{(l)}$ for $\Delta=0$ can be rewritten in terms of flux derivatives of the eigenenergies up to order $l$ only. Here we rigorously prove that there is an analogous expression for the interacting case using the Kawakami--Fujimoto~\cite{FujimotoKawakami} approach.

We start from the general formula for the non-linear Drude weight $D^{(3)}$ and reexpress the probability distribution $p_{i}=\exp(-\beta E_{i})/Z$ in terms of TBA distribution functions $p_{i} \to\rho_{\alpha}(\theta)= \rho^T_{\alpha}(\theta)  \vartheta_{\alpha}(\theta)$ leading to
\begin{eqnarray}
\label{Drude2appendix}
D^{(3)} &=& N^3\sum_{i} p_{i} \left.\frac{\partial^{4}E_{i}}{\partial \phi^{4}}\right|_{\phi=0}
\to N^4\sum_{\alpha} \int d\theta \, \rho_{\alpha}(\theta) \left.\frac{\partial^{4}E_{\alpha}(\theta)}{\partial \phi^{4}}\right|_{\phi=0}\,.
\end{eqnarray}

The flux corrections to the energy are obtained by a Taylor expansion in the finite size corrections to the thermodynamic rapidities. Of primary interest here is the fourth order correction
\begin{eqnarray}
\label{eq:energyVal}
\frac{\partial^{4}E}{\partial\phi^4} &=  &\sum_\alpha\int d\theta\,\rho_\alpha \left( e_\alpha^{(4)}  g_{1\alpha}^4 + 12 e^{(3)} g_{1\alpha}^2  g_{2\alpha} + 12 e_\alpha^{(2)}  g_{2\alpha}^2 +24 e_\alpha^{(2)}  g_{1\alpha}  g_{3\alpha} +24 e'_\alpha  g_{4\alpha}\right).
\end{eqnarray}

Each of the terms in the integrand may be rewritten in terms of TBA quantities using the same methods as in Refs.~\cite{FujimotoKawakami,zotos,benz_finite_2005}. Below we explicitly show how the terms may be recast through the dressing trick making use of the relation
\begin{eqnarray}
\beta e_\alpha &=&  \ln \eta_\alpha + \sum_\delta K_{\alpha \delta} \star \ln \left( 1 + \eta^{-1}_\delta\right),
\end{eqnarray}
which is valid at zero flux. We use $ \ln \eta_\alpha = \xi_\alpha$, $\vartheta_\alpha = \frac{1}{1+\eta_\alpha}$,  $\Lambda^{(n)}_\alpha = \vartheta_\alpha \partial^n_\theta \ln \eta_\alpha +\partial^n_\theta \ln \left(1+\eta^{-1}_\alpha\right) $, and it should be noted that $\Lambda^{(1)}_\alpha = 0$. To facilitate simplification, consider a function $f(\{g_i\})$ that is a product of $g$-functions. At infinite temperatures the terms have the particularly simple form
\begin{eqnarray}
\sum_\alpha \int d\theta && f(\{g_i\}) \rho^T_\alpha \vartheta_\alpha \beta e^{(n)}_\alpha = \sum_\alpha \int d\theta f(\{g_i\}) \rho^T_\alpha \left[ \vartheta_\alpha \partial_\theta^n \ln \eta_\alpha + \sum_\delta \vartheta_\alpha K_{\alpha \delta} \star \partial^n_\theta \ln \left(1 + \eta^{-1}_\delta\right) \right]\nn\\
&&= \sum_\alpha \int d\theta f(\{g_i\}) \rho^T_\alpha  \Lambda^{(n)}_\alpha -\sum_{\delta,\alpha} \int d\theta  \left[ f(\{g_i\}) \rho^T_\alpha \star \left(1 - \vartheta_\alpha K_{\alpha \delta}\right) \right]  \partial_\theta^n \ln \left(1 + \eta^{-1}_\delta\right) \, ,\nn\\
&&= \sum_\alpha \int d\theta f(\{g_i\}) \rho^T_\alpha  \Lambda^{(n)}_\alpha -(-1)^{n}\sum_{\delta,\alpha} \int d\theta  \left[ \partial_\mu^{n-1}\left(f(\{g_i\}) \rho^T_\alpha\right) \star \left(1 - \vartheta_\alpha K_{\alpha \delta}\right) \right] \vartheta_\delta\partial_\theta\xi_\delta +\mathcal{O}(\beta),
\end{eqnarray}
with the note that $\partial_\mu \vartheta \sim \mathcal{O}(\beta)$ is used in the last line.

From this relation and using $\rho_\alpha = \rho^T_\alpha \vartheta_\alpha$ it is straightforward to write out the terms appearing in the energy corrections term by term
\begin{eqnarray}
\label{gfcts2}
\int d\theta \, \beta e'_\alpha g_{4\alpha} \rho_\alpha  
&=&\sum_\delta \int d\theta \,  \left[\rho^T_{\delta} g_{4\delta} \star \left(1 - \vartheta_\delta K_{\delta \alpha}\right)\right] \vartheta_\alpha \partial_\theta \xi_\alpha \nn\\
\int d\theta \beta e_\alpha^{(2)}g_{1\alpha}g_{3\alpha} \rho_\alpha &=& \int d\theta g_{1\alpha}g_{3\alpha}\rho_\alpha^T \Lambda^{(2)}_\alpha - \sum_\delta \int d\theta \left[\partial_\mu\left(\rho^T_\delta g_{1\delta} g_{3\delta}\right) \star \left(1 - \vartheta_\delta K_{\delta \alpha}\right)\right] \vartheta_\alpha \partial_\theta \xi_\alpha,\nn\\
\int d\theta  \beta e^{(2)}_\alpha g^2_{2\alpha} \rho_\alpha &=& \int d\theta g_{2\alpha}^2 \rho_\alpha^T \Lambda^{(2)}_\alpha - \sum_\delta \int d\theta \left[\partial_\mu\left(\rho^T_\delta g_{2\delta}^2\right) \star \left(1 - \vartheta_\delta K_{\delta \alpha}\right)\right] \vartheta_\alpha \partial_\theta \xi_\alpha,\nn\\
\int d\theta  \beta e^{(3)}_\alpha g^2_{1\alpha}g_{2\alpha} \rho_\alpha &=& \int d\theta g^2_{1\alpha}g_{2\alpha}  \rho_\alpha^T \Lambda^{(3)}_\alpha +\sum_\delta  \int d\theta \left[\partial_\mu^2\left(\rho^T_\delta g^2_{1\delta}g_{2\delta} \right) \star \left(1 - \vartheta_\delta K_{\delta \alpha}\right)\right] \vartheta_\alpha \partial_\theta \xi_\alpha,\nn\\
\int d\theta  \beta e^{(4)}_\alpha g^4_{1\alpha} \rho_\alpha &=& \int d\theta g^4_{1\alpha} \rho_\alpha^T \Lambda^{(4)}_\alpha -\sum_\delta  \int d\theta \left[\partial_\mu^3\left(\rho^T_\delta g^4_{1\delta} \right) \star \left(1 - \vartheta_\delta K_{\delta \alpha}\right)\right] \vartheta_\alpha \partial_\theta \xi_\alpha .
\end{eqnarray}

An additional identity between infinite temperature finite-size corrections follows directly from Eq.~\eqref{eq:g_functions} 
\begin{eqnarray}
\rho^T_\alpha g_{4 \alpha} = \partial_\theta \left( \rho^T_\alpha g_{1\alpha} g_{3\alpha} \right) +\frac{1}{2}\partial_\theta \left( \rho^T_\alpha g^2_{2\alpha} \right) - \frac{1}{2} \partial_\theta^2 \left( \rho^T_\alpha g_{1\alpha}^2 g_{2\alpha} \right) + \frac{1}{24}\partial_\theta^3 \left( \rho^T_\alpha g_{1\alpha}^4 \right).
\end{eqnarray}
Here it should be noted that $\rho^T_\alpha=m(\delta_{\alpha,m}+\delta_{\alpha,m-1})/(4\pi g_{1\alpha})$, see Eq.~\eqref{gfcts}.

After combining all these relations and using them in Eq.~\eqref{eq:energyVal}, the $g_4$ function and all the terms which contain a convolution in Eq.~\eqref{gfcts2} are found to exactly cancel. Consequently, inserting the infinite-temperature relation $\partial_\theta \xi_\alpha = \beta\partial_\theta\dr{\varepsilon}_\alpha$ leads to the non-linear Drude weight expressed in terms of the dressed energies and $g$-functions as
\begin{eqnarray}
 D^{(3)}&=&\beta \sum_\alpha \int  d\theta \rho^T_\alpha \vartheta_\alpha (1-\vartheta_\alpha) \left\{  (\partial_\theta \dr\varepsilon_\alpha)^2 \left[24g_{1\alpha}g_{3\alpha}+12 g_{2\alpha}^2\right] + 36(\partial_\theta \dr \varepsilon_\alpha)(\partial^2_\theta \dr\varepsilon_\alpha) g_{2\alpha}g_{1\alpha}^2\right\}\nn\\
 &&+\beta \sum_\alpha \int  d\theta \rho^T_\alpha \vartheta_\alpha (1-\vartheta_\alpha) \left\{ 3 (\partial_\theta^2 \dr\varepsilon_\alpha)^2 + 4 (\partial_\theta^3 \dr\varepsilon_\alpha)(\partial_\theta \dr\varepsilon_\alpha) \right\}g_{1\alpha}^4 +\mathcal{O}(\beta^2).
\end{eqnarray}
This is identical to the formula which we obtained in the main text by a less stringent derivation which was based on the free fermion result combined with an appropriate replacement of bare with dressed quantities.

We can now insert the $g$-functions from Eq.~\eqref{gfcts} and obtain an integral that can be evaluated
\begin{eqnarray}
D^{(3)} &=&\beta\sum_\alpha \frac{ \dr{q}_\alpha^4\sin^3\gamma}{2\pi\gamma^3\sigma_\alpha}  \vartheta_\alpha(1-\vartheta_\alpha) \int d\theta \left\{ \frac{4(\partial_\theta \dr \varepsilon_\alpha)^2 }{\dr{\varepsilon_\alpha}} \partial_\theta\left( \frac{1}{\dr{\varepsilon}_\alpha} \partial_\theta \left(\frac{1}{\dr{\varepsilon}_\alpha} \right)\right)+ \frac{3(\partial_\theta \dr \varepsilon_\alpha)^2 }{\dr{\varepsilon}_\alpha}\left(\partial_\theta \frac{1}{\dr{\varepsilon}_\alpha} \right)^2 \right\} \nn \\ 
&+&\beta\sum_\alpha\frac{\dr{q}_\alpha^4\sin^3\gamma}{2\pi\gamma^3\sigma_\alpha}  \vartheta_\alpha(1-\vartheta_\alpha) \int d\theta  \left\{  \frac{18(\partial_\theta^2 \dr\varepsilon_\alpha) (\partial_\theta \dr\varepsilon_\alpha)}{\dr{\varepsilon}^2_\alpha} \partial_\theta\left(\frac{1}{\dr{\varepsilon}_\alpha}\right)+\frac{3 (\partial_\theta^2 \dr \varepsilon_\alpha)^2 + 4 (\partial_\theta^3 \dr \varepsilon_\alpha) (\partial_\theta \dr \varepsilon_\alpha)    }{\dr{\varepsilon}^3_\alpha}\right\} + \mathcal{O}\left(\beta^2\right).
\end{eqnarray}
The presence of $\dr{q}_\alpha$ makes it so that only the last two strings have non-zero contributions to the infinite temperature non-linear Drude weight with an overall sign $\sigma_\alpha = \pm 1$. Notably, both of the last two strings contribute identically, so the sum over string types may be carried out trivially. After evaluating the sum we write everything in terms of the second last string, so drop the string index, to find
\begin{eqnarray}
\label{Eq18}
D^{(3)} =&&2 \sigma \frac{\beta \dr{q}^4\sin^3(\gamma)}{2\pi\gamma^3}  \vartheta(1-\vartheta) \int d\theta \left\{ \frac{4(\partial_\theta \dr \varepsilon)^2 }{\dr{\varepsilon}} \partial_\theta\left( \frac{1}{\dr{\varepsilon}} \partial_\theta \left(\frac{1}{\dr{\varepsilon}} \right)\right)+ \frac{3(\partial_\theta \dr \varepsilon)^2 }{\dr{\varepsilon}}\left(\partial_\theta \frac{1}{\dr{\varepsilon}} \right)^2 \right\} \nn \\ 
&+&2\sigma\frac{\beta\dr{q}^4\sin^3(\gamma)}{2\pi\gamma^3}  \vartheta(1-\vartheta) \int d\theta  \left\{  \frac{18(\partial_\theta^2 \dr\varepsilon) (\partial_\theta \dr\varepsilon)}{\dr{\varepsilon}^2} \partial_\theta\left(\frac{1}{\dr{\varepsilon}}\right)+\frac{3 (\partial_\theta^2 \dr \varepsilon)^2 + 4 (\partial_\theta^3 \dr \varepsilon) (\partial_\theta \dr \varepsilon)    }{\dr{\varepsilon}^3}\right\} + \mathcal{O}\left(\beta^2\right).
\end{eqnarray}

This formula may be simplified further by grouping the derivatives together leading to
\begin{eqnarray}
\label{eq:infiniteAsymp}
D^{(3)} &\sim&\beta\frac{\sin^3\gamma }{ \pi  \gamma^3} \dr{q}^4 \sigma \vartheta(1-\vartheta)\left[3 \frac{\partial_\theta \dr{\varepsilon}}{\dr{\varepsilon}^2} \partial_\theta \left(\frac{\partial_\theta \dr{\varepsilon}}{\dr{\varepsilon}}\right) \bigg\rvert_{\pm \infty}+\int d\theta \,  \frac{\partial_\theta \dr{\varepsilon}}{\dr{\varepsilon}}  \, \partial_\theta\left(\frac{1}{\dr{\varepsilon}} \partial_\theta\left(\frac{\partial_\theta \dr{\varepsilon}}{\dr{\varepsilon}} \right) \right)  \right]\,.
\end{eqnarray}
The expression in Eq.~\eqref{eq:infiniteAsymp} is remarkably similar to that obtained in the $\Delta>1$ case discussed in Ref.~\cite{fava_hydrodynamic_2021} albeit with an additional integral of a total derivative that does not appear in their expression. 

\section{Appendix C: Entropy formula}
A less rigorous procedure which does lead to a compact formula for all non-linear Drude weights can be carried out based on the following observation. The Drude weight is not a bulk thermodynamic quantity, i.e.~it cannot be obtained from the inner energy or the entropy of the system. However, it is closely related to a derivative of the inner energy. In the free fermion case, the non-linear Drude weight is given by $D^{(l)}=N^l\sum_k n_k \varepsilon_k^{(l+1)}(\phi)$ which can be viewed as the flux derivative of a modified energy where only the dispersion $\varepsilon_k(\phi)$ depends on the flux $\phi$ while the occupation number $n_k$ is considered as flux independent. Similarly, in the interacting case we observe that the Drude weight formulas can be viewed as stemming from modified thermodynamic quantities where only the rapidities entering the $\eta$-functions depend on the flux while the distribution function $\rho^T(\theta)$ and $\vartheta(\theta)$ are always evaluated at the thermodynamic limit rapidity. Using this ,recipe' we can define a modified energy and a modified entropy as
\begin{eqnarray}
\label{mod_fcts}
\tilde{e}&=&\sum_\alpha\int d\theta\, \rho^T_\alpha(\theta)\vartheta_\alpha(\theta)E(\theta^L_\alpha(\phi)) \nonumber \\
\tilde{s}&=& \sum_\alpha\int d\theta\, \rho^T_\alpha(\theta)\left[\ln\{1+\eta_\alpha^{-1}(\theta^L_\alpha(\phi))\}+\vartheta_\alpha(\theta)\ln\eta_\alpha(\theta^L_\alpha(\phi))\right] \, .
\end{eqnarray}
If we assume, furthermore, that the flux derivatives of the free energy $\tilde{f}=\tilde{e}-T\tilde{s}$ vanish, then we obtain the following modified entropy formula for the non-linear Drude weight
\begin{eqnarray}
\label{compact_formula}
\beta D^{(l)} &=&N^{l+1} \left. \partial_\phi^{l+1} \beta\int d\theta \rho(\theta) E(\theta^L)\right|_{\phi=0} =\left. N^{l+1} \int d\theta  \rho^T(\theta)\left[ \partial_\phi^{l+1} \ln \left(1+\eta^{-1}\right)(\theta^L) + \vartheta(\theta) \partial_\phi^{l+1}\ln \eta(\theta^L)\right]\right|_{\phi=0} \!\!\!\!\! .
\end{eqnarray}
The flux derivatives are evaluated as before by expanding $\theta^L$ around the thermodynamic limit rapidity. This is best illustrated by an example: It is straightforward to read off the linear Drude weight $(l=1)$ as
\begin{eqnarray}
D^{(1)} &=& \frac{1}{\beta} \int d\theta \rho^T(\theta)\left[  \partial_\phi^2 \ln \left(1+\eta^{-1}\right)(\theta^L) + \vartheta(\theta)\partial_\phi^2\ln \eta(\theta^L)\right]\nn\\
&=& \frac{1}{\beta} \int d\theta \rho^T(\theta)\left[\vartheta (1-\vartheta) \left[ \partial_\phi \ln \eta(\theta^L)\right]^2\right]
= \frac{1}{\beta} \int d\theta \rho^T \vartheta (1-\vartheta) g_1^2\left[ \partial_\theta \ln \eta(\theta)\right]^2 \, .
\end{eqnarray}
Non-linear Drude weights $D^{(l)}$ with $l>1$ can likewise be computed by considering higher order flux derivatives. Possible non-analytic behaviour at certain anisotropies must be taken into consideration~\cite{tanikawa_fine_2021}, however as long as the anisotropy is in the analytic region then Eq.~\eqref{compact_formula} is applicable and directly yields a compact formula containing only $g$-functions up to order $l$.

We want to stress that the formula \eqref{compact_formula} is again based on an analogy with the free fermion case and that we do not have a strict proof. A rigorous proof of the non-linear Drude weight formula for $D^{(3)}$ based on the Kawakami-Fukimoto approach was presented in the previous section.

\section{Appendix D: High-temperature Analytical Result}
\label{sec:fermiHighT}
In order to carry out the explicit high-temperature asymptotic calculation, the underlying $T$ and $Y$-systems~\cite{kuniba_t-systems_2011} are used to determine the Fermi weight corresponding to the second last string. The setup for this analysis is carried out in more detail in~\cite{urichuk_spin_2019,urichuk_analytical_2021} for the high and low-temperature asymptotics, respectively. At the heart of this analysis are the following relations 
\begin{eqnarray}
\label{eq:fermiWeightEval}
m\vartheta(\theta) &=&\sum_{j=1}^{\bar \mu} \frac{\phi^{\ii (s+2j)}(\theta)}{Q^{\ii (s+2j-1)}(\theta)Q^{\ii (s+2j+1)}(\theta)}, \\
\phi(\theta) &=& \left(\sinh\left[\frac{\gamma(\theta+\ii u+\ii)}{2} \right]\sinh\left[\frac{\gamma(\theta-\ii u-\ii)}{2} \right]\right)^{N/2}, \nn \\
Q(\theta)&=&\prod_{ \{\lambda_j\}}\sinh\left[\frac{\gamma (\theta-\lambda_j)}{2}\right].\nn
\end{eqnarray}
Here $\phi^{\ii \tau}(\theta)=\phi(\theta+\ii \tau)$, $\lambda_j$ are Bethe roots, $u=-\frac{ \beta \sin\gamma}{\gamma N}$ with $N$ being the Trotter number, and the shift $s$ is connected to the Takahasi-Suzuki integers
\begin{equation}
\label{eq:TSshift}
s+1 = 2m - 2\bar \mu + \frac{(-1)^{1+c}}{n}-p_0,
\end{equation}
which depend on the rational anisotropy $\gamma/\pi = \frac{n}{m}$ written as a continued fraction of length $c$, rational numbers $p_0=\frac{\pi}{\gamma}$, and the last string length $\bar \mu$ that is related to the second last string length $\mu$ as $m=\bar \mu + \mu$~\cite{kuniba_continued_1998}. 

In Ref.~\cite{urichuk_spin_2019} we have previously shown that to  leading order in inverse temperature $\beta$ the Fermi weight $\vartheta = 1/(1+\eta)$ corresponding to the second last string is given by
\begin{eqnarray}
(1+ \eta) \approx \frac{m}{\bar \mu} \left(1 + \frac{\ii  \beta\sin\gamma}{4\bar \mu} \left[\coth\left(\frac{\gamma}{2}(\theta + \ii(s+1))\right)-\coth\left(\frac{\gamma}{2}(\theta + \ii(s+1)+2\ii \bar\mu)\right)\right]\right).
\end{eqnarray}
To simplify this relation, we require a single additional identity for the TS-integers. First, we note that $\bar \mu = y_{m_c-1}$. Then, in the notation of Ref.~\cite{kuniba_continued_1998}, the relation 
\begin{eqnarray}
\bar \mu \gamma = \pi z_{c-1} + (-1)^{1+c} \gamma p_c = \pi z_{c-1}  + \frac{\pi (-1)^{c+1}}{m}
\end{eqnarray}
holds with $z_{c-1}\in \mathbb{Z}_{\geq 0}$. That $p_c = 1/n$ was proven in Ref.~\cite{urichuk_spin_2019}. With these relations the Fermi weight may be rewritten as 
\begin{eqnarray}
(1+ \eta) \approx \frac{m}{\bar \mu} \left(1 +(-1)^{c+1} \frac{ \beta }{4\bar \mu}\frac{\sin\gamma\sin\left( \frac{\pi}{m} \right)}{\cosh\left(\frac{\gamma \theta}{2}- \frac{\ii\pi }{2m}  \right)\cosh\left(\frac{\gamma \theta}{2} + \frac{\ii\pi}{2m}  \right)} \right)
\end{eqnarray}
valid for any anisotropy. Before proceeding further, note that $\sigma_j = \text{sign}(q_j)$ with $q_j$ being a rational number defined in~\cite{takahashi_thermodynamics_1999} along with $m_k = \sum_{i=1}^k \nu_i$, where the $\nu_i$ are positive integers in the continued fraction $\pi/\gamma = [\nu_1, \dots, \nu_c]$. With these definitions $m_c - m_{c-1} = \nu_c = \frac{p_{c-1}}{p_c}$ and consequently the $q$-number corresponding to the second last string is
\begin{eqnarray}
q_{m_c-1} = (-1)^{c-1}\left(p_{c-1} - (m_c - m_{c-1}-1) p_c\right) = (-1)^{c-1}p_c = \frac{(-1)^{c-1}}{n},
\end{eqnarray}
this then implies that $\sigma = (-1)^{c-1}$.

We note that in Eqs.~(\ref{Eq18},\ref{eq:infiniteAsymp}) we already have an explicit factor of $\beta$ so to obtain the leading asymptotics only the $\mathcal{O}(\beta^0)$ contributions from the other terms are needed. In particular, $\dr{q}^4 \vartheta(1-\vartheta)\sim \left(\frac{m}{2}\right)^4\frac{(m-\bar\mu) \bar\mu}{m^2}$ and, using $\dr{\varepsilon}=\partial_\beta \ln \eta = \frac{1}{(1+\eta)-1} \partial_\beta(1+\eta)$,
\begin{eqnarray}
\label{dressedEAppendix}
\dr{\varepsilon} \sim (-1)^{c+1}\frac{m}{ 4\bar \mu (m-\bar \mu)} \frac{ \sin\gamma \sin\left(\frac{\pi}{m}\right)}{\cosh\left(\frac{\gamma \theta}{2}- \frac{\ii\pi }{2m}  \right)\cosh\left(\frac{\gamma \theta}{2} + \frac{\ii\pi}{2m}  \right)}.
\end{eqnarray}
It is easy to check that for the simple roots of unity case, $\gamma=\pi/m$, this agrees with Eq.~(12) in the main text.
This formula for $\dr\varepsilon$ can now be inserted into Eq.~\eqref{Eq18} or Eq.~\eqref{eq:infiniteAsymp}, followed by an evaluation of the integral. Doing so, we obtain the asymptotics
\begin{eqnarray}
\label{main}
TD^{(3)}= -\underbrace{\sigma (-1)^{c+1}}_{=1}  \frac{m^4 }{8}\underbrace{\frac{((m-\bar \mu)\bar \mu)^2}{m^4}}_{\{\vartheta(1-\vartheta)\}^2}\frac{\sin^2\gamma}{ \sin^2\left(\frac{\pi}{m}\right)} \left[ 1 - \frac{7m}{2\pi} \sin\left(\frac{2\pi}{m}\right)\right]\,.
\end{eqnarray}

It is worth re-emphasizing that in contrast to the linear case this result depends on the Fermi weight $\vartheta$ which is related to the last string length $\bar \mu$ of the bound states transporting the current by $\vartheta=\frac{\bar \mu}{m}$.

\section{Appendix E: Free fermions}
\label{fF}
\begin{figure}
    \centering
    \includegraphics[width=0.5\textwidth]{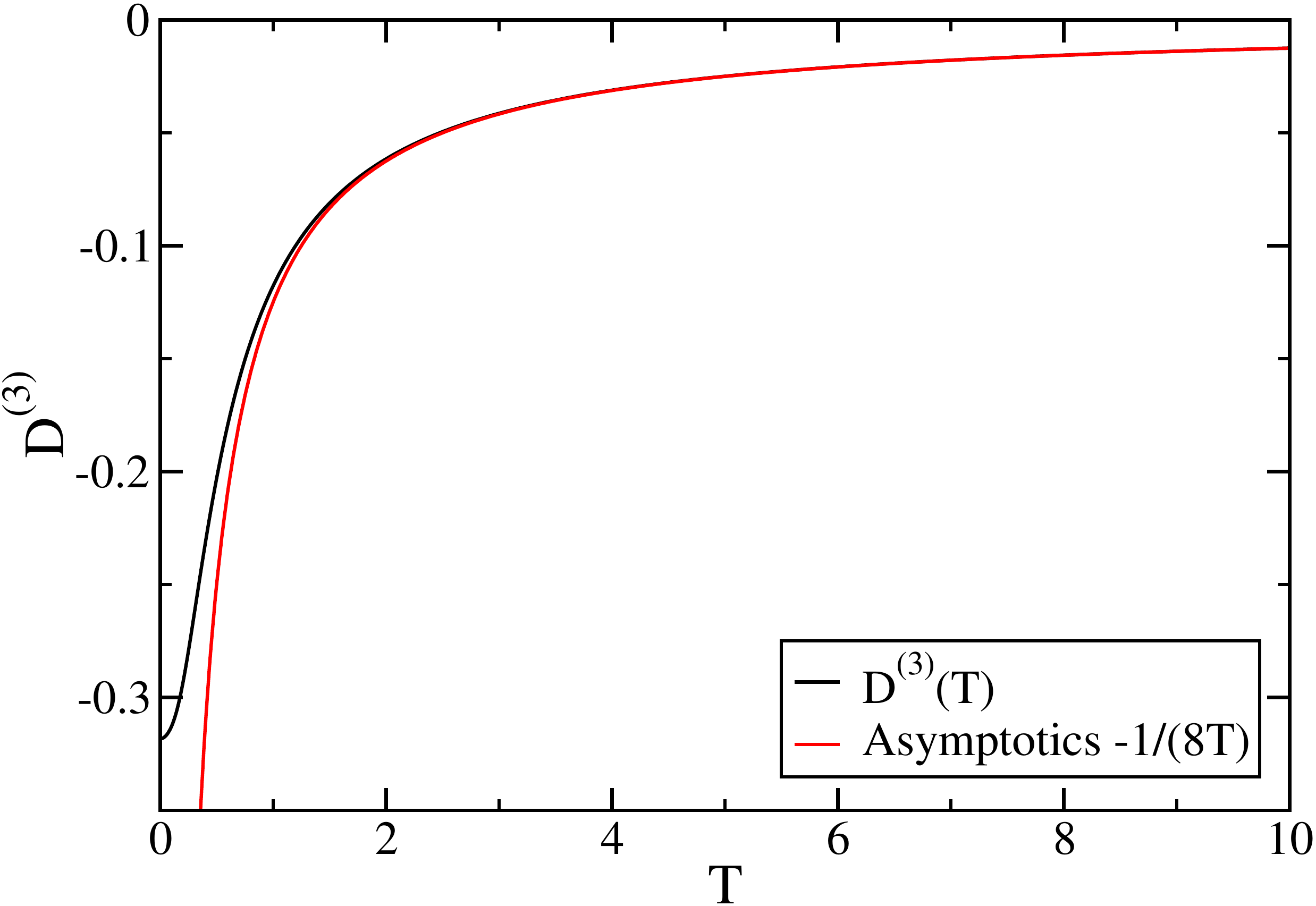}
    \caption{The Drude weight $D^{(3)}$ for $\Delta=0$ as a function of temperature. Also shown is the high-temperature asymptotics.}
    \label{FigD0}
\end{figure}
As a further check of our main result, we consider the case $\Delta=0$ ($n=1,m=2$) where Eq.~\eqref{main} reduces to $D^{(3)}=-\beta/8$. The XXZ spin-1/2 chain can be mapped onto a model of interacting spinless fermions using the Jordan-Wigner transformation
\begin{equation}
    \label{JW}
S^+_j\to(-1)^j c_j^\dagger \e^{\ii\pi\varphi_j},\;  S^-_j\to(-1)^j c_j \e^{-\ii\pi\varphi_j}, \; S^z_j\to n_j -\frac{1}{2}   
\end{equation}
where $\varphi_j=\sum_{\ell=1}^{j-1} n_l$ resulting in 
\begin{equation}
    \label{HamF}
H=\sum_j\left\{-\frac{1}{2}\left(\e^{-\ii\phi/N}c_j^\dagger c_{j+1} +h.c.\right) + \Delta \left(n_j-\frac{1}{2}\right)\left(n_{j+1}-\frac{1}{2}\right)\right\} \, .    
\end{equation}
Here $c_j$ ($c_j^\dagger$) are annihilation (creation) operators at site $j$ and $n_j=c_j^\dagger c_j$. For $\Delta=0$ the model describes non-interacting spinless fermions and can be diagonalized by a Fourier transform. This leads to $H=\frac{1}{N}\sum_k\varepsilon_k c_k^\dagger c_k$ with dispersion $\varepsilon_k=-\cos(k+\phi/N)$.

For the non-linear Drude weight $D^{(3)}$ in the non-interacting case we have found the following simple expression
\begin{equation}
\label{Drude_D0App}
D^{(3)} = N^3\sum_k n_k\varepsilon_k^{(4)} \, .
\end{equation}
Here $n_k=1/(\exp(\beta\varepsilon_k)+1)$ is the Fermi-Dirac distribution. Noting that every flux derivative yields a factor $1/N$ and replacing $\frac{1}{N}\sum_k\to\frac{1}{2\pi}\int_{-\pi}^\pi dk$ we can now perform the thermodynamic limit and obtain 
\begin{equation}
    \label{Drude_D0_2}
D^{(3)} = -\frac{1}{2\pi}\int_{-\pi}^\pi dk\, \frac{\cos k}{\e^{-\beta\cos k} +1}  ,
\end{equation}
a formula which can be evaluated at all temperatures. More generally, we see from Eqs.~\eqref{Drude_D0App},\eqref{Drude_D0_2} that $D^{(l)}=N^l\sum_k n_k\varepsilon_k^{(l+1)}$ is zero for $l$ even and $D^{(l)}=(-1)^{(l-1)/2}D^{(1)}$ for $l$ odd. For $\beta\to 0$ we can Taylor expand \eqref{Drude_D0_2} and confirm that
\begin{equation}
    \label{DT0}
D^{(3)}_{T\to \infty} = -\frac{\beta}{8\pi}\int_{-\pi}^\pi \cos^2k = -\frac{\beta}{8} \, .
\end{equation}
For $T=0$, the integral \eqref{Drude_D0_2} is reduced to $D^{(3)}_{T=0}=-\frac{1}{2\pi}\int_{-\pi/2}^{\pi/2}\cos k = -1/\pi$. The temperature dependence and the high-temperature asymptotics are shown in Fig.~\ref{FigD0}.

\section{Appendix F: Exact Diagonalization}
\begin{figure}
    \centering
    \includegraphics[width=0.7\textwidth]{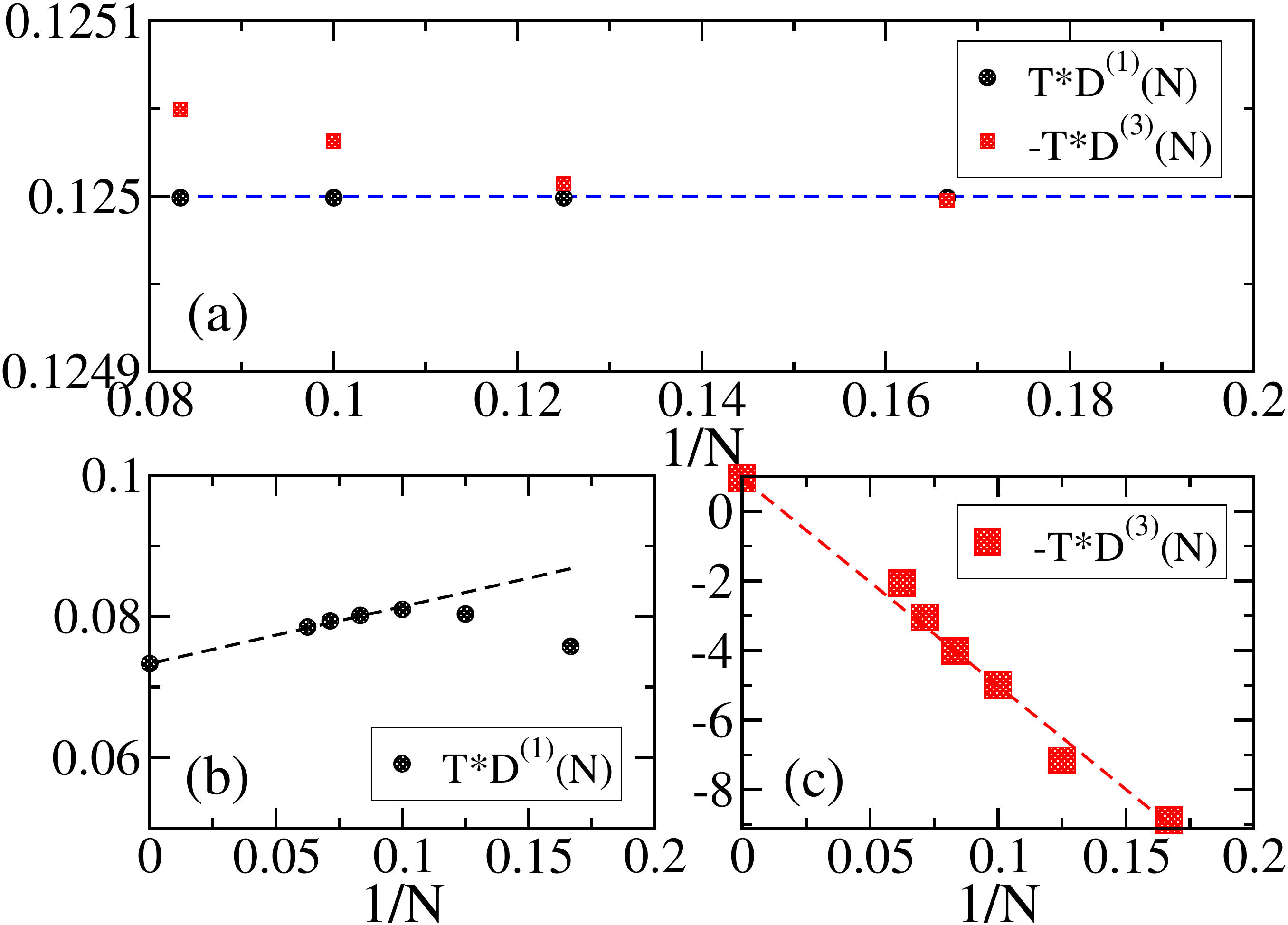}
    \caption{(a) $TD^{(1)}$ and $TD^{(3)}$ from exact diagonalization for finite systems of length $N$ and anisotropy $\Delta=0$. The dashed line denotes the thermodynamic limit value $1/8$. (b) $TD^{(1)}$ for $\Delta=-1/2$. (c) $-TD^{(3)}$ for $\Delta=-1/2$. Note the large finite size corrections. The dashed lines in (b,c) denote linear fits with the $1/N=0$ value fixed by the analytical result, Eq.~(12) in the main text.}
    \label{Fig_ED}
\end{figure}
As an additional check of our analytic formula \eqref{main}, we calculate $D^{(3)}$ for finite system sizes evaluating Eq.~(2) in the main text based on exact diagonalizations of the Hamiltonian. Apart from being restricted to finite system sizes, another problem is that first, second, and third derivatives w.r.t.~the flux have to be evaluated numerically. Here we find that the following method gives the most stable results: we calculate the energies for $50-100$ different small flux values. We then fit each energy value $E_i(\phi)$ by a polynomial in $\phi$. We start with a second order polynomial and then check if a third order polynomial improves the standard deviation by at least one order of magnitude. Otherwise, we keep the second order polynomial fit and set the higher order terms to zero. We continue this process up to at most a fourth order polynomial fit. From the parameters of the polynomial fits we then determine the flux derivatives.

To show that this method can give reliable results, we start with the free fermion case. For comparison, we also show the linear Drude weight which can be computed by
\begin{equation}
    \label{D1}
D^{(1)} = \frac{N}{Z}\sum_i\e^{-\beta E_i}\frac{\partial^2 E_i}{\partial\phi^2} = \frac{N\beta}{Z}\sum_i \e^{-\beta E_i}\left(\frac{\partial E_i}{\partial\phi}\right)^2 \, .
\end{equation}
The numerical results for various system sizes are shown in Fig.~\ref{Fig_ED}(a). For both $D^{(1)}$ and $D^{(3)}$ we find that the finite size corrections at $\Delta=0$ are very small. While calculating flux derivatives up to third order instead of just the first order derivative does increase the error for $D^{(3)}$ as compared to $D^{(1)}$, we note that all the data still only have a relative deviation from the thermodynamic limit result of less than $10^{-5}$.

Next, we show numerical results for the interacting case $\Delta=\cos(2\pi/3)=-0.5$ in Fig.~\ref{Fig_ED}(b,c). Instead of trying to extrapolate the numerical data for $N\to\infty$ we have a more modest goal here and want to simply show that the finite-size scaling of the data is consistent with the analytical result, Eq.~(12) in the main text. For the linear Drude weight we find that the finite-size corrections are still relatively small; all data points are within $10^{-2}$ of the thermodynamic limit result, see Fig.~\ref{Fig_ED}(b). In contrast, the finite-size corrections for $D^{(3)}$ are large and the result for the smallest system size $N=6$ deviates by about a factor $10$ from the thermodynamic limit result and also has the opposite sign, see Fig.~\ref{Fig_ED}(c). Nevertheless, the linear fit shows that all the data are consistent with the analytical result in the thermodynamic limit.
 
\end{widetext}


%

\end{document}